\documentclass[
draft            % use draft while you are working on the paper
  ]
  {aipproc}

\layoutstyle{8x11single}

\usepackage{amssymb}
\usepackage[varg]{txfonts}

\newcommand{\be}{\begin{equation}}
\newcommand{\ee}{\end{equation}}
\newcommand{\bea}{\begin{eqnarray}}
\newcommand{\eea}{\end{eqnarray}}

\newcommand{\6}{\partial }

\newcommand{\R}{r_{\rm KK}}
\newcommand{\RSS}{R_{\rm D4}}
\newcommand{\MKK}{M_{\rm KK}}

\newcommand{\gYM}{g_{\rm YM}}
\newcommand{\Nc}{N_{c}}
\newcommand{\ls}{l_{s}}
\newcommand{\gs}{g_{s}}

\newcommand{\Tr}{{\rm Tr}\,}

\begin{document}

\title{Top-down Holographic Glueball Decay Rates} 


\author{F. Br\"unner}{
  address={Institut f\"ur Theoretische Physik, Technische Universit\"at Wien,\\
        Wiedner Hauptstrasse 8-10, A-1040 Vienna, Austria}
}

\author{D. Parganlija}{
%  address={<common address for author2 and author3>}
}

\author{\underline{A. Rebhan}}{
%  address={<common address for author2 and author3>}
%  ,altaddress={<author1 address>} % additional visiting address
}

\begin{abstract}
We present new results on the decay patterns of scalar and tensor glueballs in the top-down holographic Witten-Sakai-Sugimoto model. This model, which has only one free dimensionless parameter, gives semi-quantitative predictions for the vector meson spectrum, their decay widths, and also a gluon condensate in agreement with SVZ sum rules. The holographic predictions for scalar glueball decay rates are compared with experimental data for the widely discussed gluon candidates $f_0(1500)$ and $f_0(1710)$.
\end{abstract}

\maketitle

%%%%%%%%%%%%%%%%%%%%%%%%%%%%%%%%%%%%%%%%%%%%
%% MAINMATTER
%%%%%%%%%%%%%%%%%%%%%%%%%%%%%%%%%%%%%%%%%%%%

\section{Introduction}

Glueballs, color-neutral bound states of gluons, are the only physical states available
in pure Yang-Mills theory and are also expected to show up in the meson
spectrum of quantum chromodynamics (QCD) \cite{Fritzsch:1972jv,Fritzsch:1975tx,Jaffe:1975fd}. The mass of the lowest glueball state
with $J^{PC}=0^{++}$ is predicted by lattice gauge theory \cite{Morningstar:1999rf,Chen:2005mg,Loan:2005ff,Gregory:2012hu}
to be in the range 1.5-1.8 GeV. However, as reviewed in Refs.~\cite{Klempt:2007cp,Crede:2008vw,Ochs:2013gi},
the identification of glueballs in the meson spectrum has remained elusive
and will be in the focus of the PANDA experiment at FAIR \cite{Lutz:2009ff,Wiedner:2011mf}.
Various scenarios have been proposed, from unobservably broad glueball resonances \cite{Ellis:1984jv,Minkowski:1998mf}
to narrow glueballs that are identified with a dominant component of one of the fairly narrow isoscalar mesons
in the mass range of the lattice results,
$f_0(1500)$ or $f_0(1710)$ \cite{Amsler:1995td,Lee:1999kv,Close:2001ga,Giacosa:2005zt,Albaladejo:2008qa,Janowski:2011gt,Janowski:2014ppa}.

QCD sum rules seem to require a broad scalar glueball \cite{Ellis:1984jv,Janowski:2014ppa}, whereas the usually
quite reliable large-$N_c$ limit would indicate narrow states.
The situation is similarly difficult in the case of tensor glueballs, which lattice simulations
predict to occur in the range 2.2-2.6 GeV, but a clear identification of a corresponding $f_2$ meson
is missing.

More information from first-principle approaches on the properties of glueballs would clearly be helpful.
Lattice gauge theory should also be able to provide information on decay rates, but
the extraction of real-time quantities is difficult, and glueballs are particularly
difficult to investigate in the presence of dynamical quarks.

\section{Holographic glueball spectrum in the Witten model} 

A new approach to study glueballs and their properties has become
available with the advent of gauge-gravity %duality and
constructions where the supersymmetry and conformal symmetry of
the underlying AdS/CFT correspondence is broken. 
The first and still most important example of a top-down holographic approach
towards (large-$N_c$) QCD is the Witten model \cite{Witten:1998zw,Aharony:1999ti} which is based on $N_c$ D4 branes in type-IIA
supergravity where one spatial dimension is compactified on a Kaluza-Klein circle
with antiperiodic boundary conditions for the fermionic gauginos.
This not only renders the gauginos massive, but through loop corrections also all adjoint
scalar matter, so that in the limit where the Kaluza-Klein mass is sent to
infinity one would arrive at pure-glue large-$N_c$ Yang-Mills theory.
Unfortunately, this limit is not accessible without leaving the realm of supergravity and
going over to a full gauge-string duality which is too difficult for being of practical use.
However, the hope is that already the supergravity approximation allows one to
study nonperturbative features of large-$N_c$ gauge theories qualitatively and
that perhaps certain quantitative results are not too far from those of actual QCD,
which is indeed frequently approximated quite well by large-$N_c$ results.

\begin{figure}[h]
  \includegraphics[height=.4\textheight]{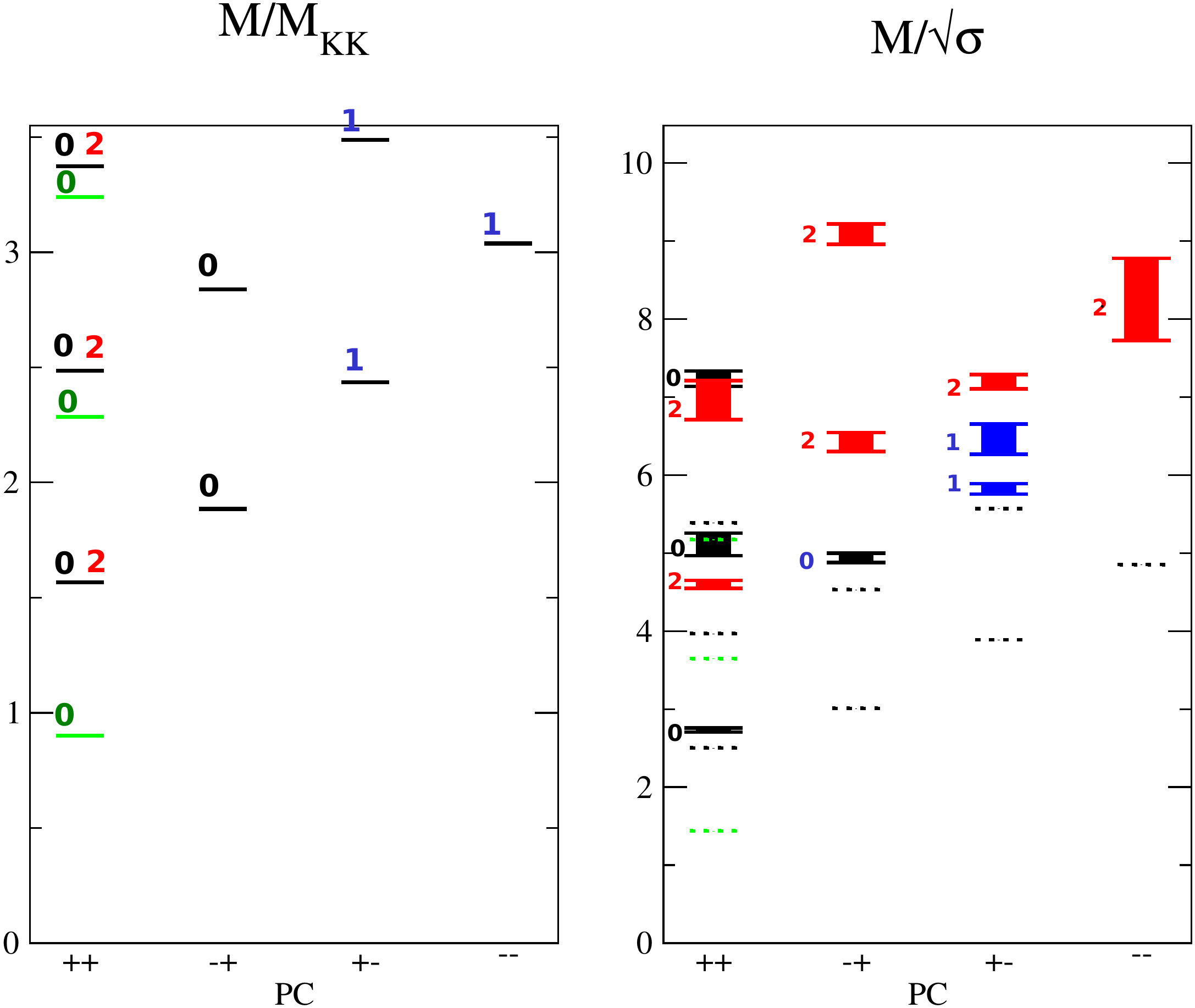}
  \caption{The holographic glueball spectrum of the Witten model (left panel, masses in units
  of $\MKK$, ``exotic'' scalar mode in green) compared to
  the recent large-$N_c$ lattice results of Ref.~\cite{Lucini:2010nv} (right panel, masses in units
  of the square root of the string tension). Dotted lines give the holographic glueball masses divided by $\sqrt{\sigma}$ from
  (\ref{sigmastring}) with $\lambda=16.63$.}
  \label{figgbspctrm}
\end{figure}

The supergravity background of the Witten model can be obtained by
another dimensional reduction, namely a supersymmetry-preserving
circle compactification of 11-dimensional supergravity with
$x^{11}\simeq x^{11}+2\pi R_{11}$, %, $R_{11}=\gs\ls$,,\quad \ls^2=\alpha'
which turns the AdS/CFT correspondence available for %so-called 
M5 branes
to a nonconformal gauge-gravity correspondence for D4 branes.
After introducing the supersymmetry-breaking circle compactification
$x^4\simeq x^4+2\pi R_4$, $R_4\equiv \MKK^{-1}$, the gravitational background is given by a
doubly Wick-rotated AdS$_7$ black hole geometry with (11-dimensional) metric
\bea
ds^2&=&\frac{r^2}{L^2}\left( f(r)dx_4^2+\eta_{\mu\nu}dx^\mu dx^\nu
+dx_{11}^2 \right)+\frac{L^2}{r^2}\frac{dr^2}{f(r)}+\frac{L^2}{4}d\Omega_4^2,\nonumber\\
&&\quad f(r)=1-\frac{\R^6}{r^6}, \quad \R=\frac{L^2}{3}\MKK,
\eea
where $L$ is the curvature radius of the asymptotic AdS$_7$ space forming a product space
with an $S^4$ of radius $L/2$.
With $x^4$ and $x^{11}$ compactified, the boundary theory at sufficiently
low energies is 4-dimensional Yang-Mills theory
with parameters
\be\label{gYMNc}
\gYM^2=\frac{g_5^2}{2\pi R_4}=2\pi\gs\ls\MKK,\quad
(L/2)^3\equiv \RSS^3=\pi \gs \Nc \ls^3.
\ee
Here $\gYM$ is the Yang-Mills coupling at the scale $\MKK$, normalized such that
the QCD coupling $\alpha_s\equiv g^2/(4\pi)=\gYM^2/(2\pi)$ (see Ref.~\cite{Rebhan:2014rxa} for a discussion
of this point).

One of the earliest applications of gauge/gravity duality has been the calculation
of the glueball spectrum from the supergravity background of the Witten model 
\cite{hep-th/9805129,Csaki:1998qr,Hashimoto:1998if,%de Mello Koch:1998qs,Csaki:1999vb,
Constable:1999gb,Brower:2000rp}. For this
one needs to determine the spectrum of fluctuations of the metric and the 3-form field
of the 11-dimensional supergravity theory. When classified
according to their symmetries in the noncompact dimensions $x^0,\ldots,x^3$ of the boundary
theory this yields towers of scalar, vector, and tensor glueballs.
This has been worked out completely in Ref.~\cite{Brower:2000rp}, with results
displayed in Fig.~\ref{figgbspctrm} and compared with recent lattice results for the glueball
spectrum in large-$N_c$ gauge theory.
Simpler (bottom-up) models \cite{BoschiFilho:2002ta,Colangelo:2007pt,Forkel:2007ru} 
typically have fewer towers of glueballs, which include scalar glueballs dual to the dilaton, and
tensor glueballs dual to the metric field.

In the Witten model, the dilaton mode of type-IIA supergravity arises as
\be
\delta\Phi=\frac{3 L^2}{4r^2}\,\delta G_{11,11}.
\ee
A scalar mode is obtained by having fluctuations of the form
\bea\label{deltaGD}
\delta G_{11,11}&=&-3\frac{r^2}{L^2}H_D(r)G_D(x),\nonumber\\
\delta G_{\mu\nu}&=&\frac{r^2}{L^2}H_D(r)\left[\eta_{\mu\nu}-\frac{\6_\mu\6_\nu}{\Box}\right]G_D(x),
\eea
which also involves the trace part of metric fluctuations.
The mass of this mode turns out to be degenerate with tensor modes from
transverse-traceless fluctuations of only $\delta G_{\mu\nu}$.

As noted first in Ref.~\cite{Constable:1999gb}, the Witten model has another scalar mode associated with 
fluctuations involving also $\delta G_{44}$, a metric fluctuation 
with components pertaining to the compactification dimension:
\bea\label{deltaGG}
&&\delta G_{44} = -\frac{r^2}{L^2}fH_G(r)G(x) \nonumber\\
&&\delta G_{\mu\nu} = \frac{r^2}{L^2}
H_G(r)\left[
\frac14 \eta_{\mu\nu} 
- \left(\frac14 + \frac{3\R^6}{5r^6-2\R^6}\right) 
\frac{\partial_\mu \partial_\nu}{M_E^2}
\right] G(x),\nonumber\\
&&\delta G_{11,11} = \frac{r^2}{L^2}\frac14
 H_G(r)G(x), \nonumber \\
&&\delta G_{rr} = -\frac{L^2}{r^2}f^{-1} \frac{3\R^6}{5r^6-2\R^6}
H_G(r) G(x) ,\nonumber\\
&&\delta G_{r\mu} = \frac{90\, r^7 \R^6}{M_E^2 L^2 (5r^6-2\R^6)^2} 
H_G(r)\partial_\mu G(x).
\eea
Ref.~ \cite{Brower:2000rp} subsequently found that this ``exotic polarization'' \cite{Constable:1999gb} has
the smallest mass of all the supergravity modes.

As displayed in Fig.~\ref{figgbspctrm},
this does make the overall spectrum of the holographic glueballs appear to qualitatively agree
with what is found in lattice gauge theory, but it leaves a certain abundance of scalar
modes in the $J^{PC}=0^{++}$ sector. Because the mass scale $\MKK$ is not fixed
in the Witten model, a more quantitative comparison with the lattice results
is not possible. However, this changes when one extends the model
to incorporate also quarks.

\section{Interactions with chiral quarks and mesons in the Sakai-Sugimoto model}

\begin{figure}[t]
\centerline{\includegraphics[width=0.4\textwidth]{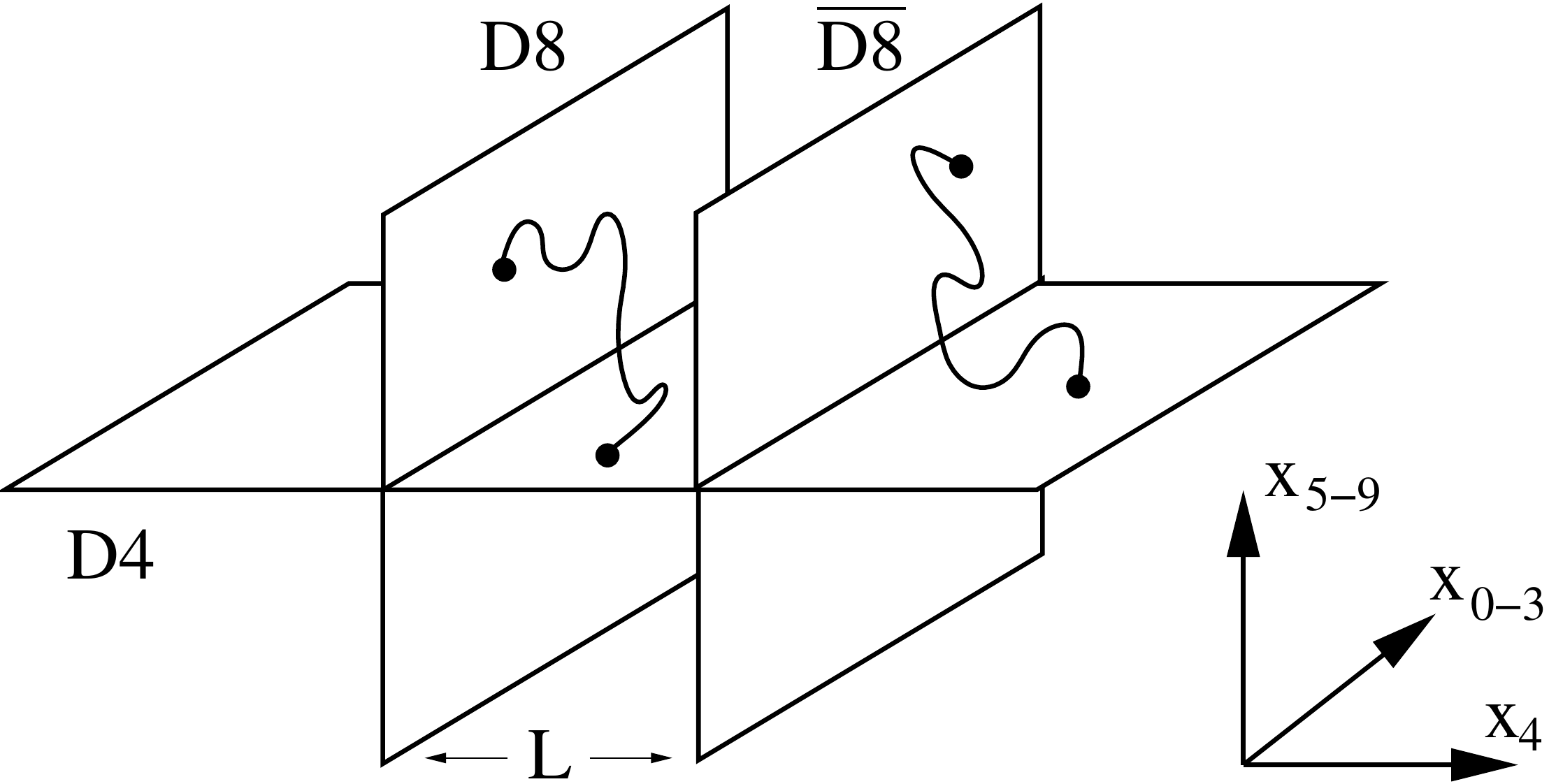}\qquad\includegraphics[width=0.35\textwidth]{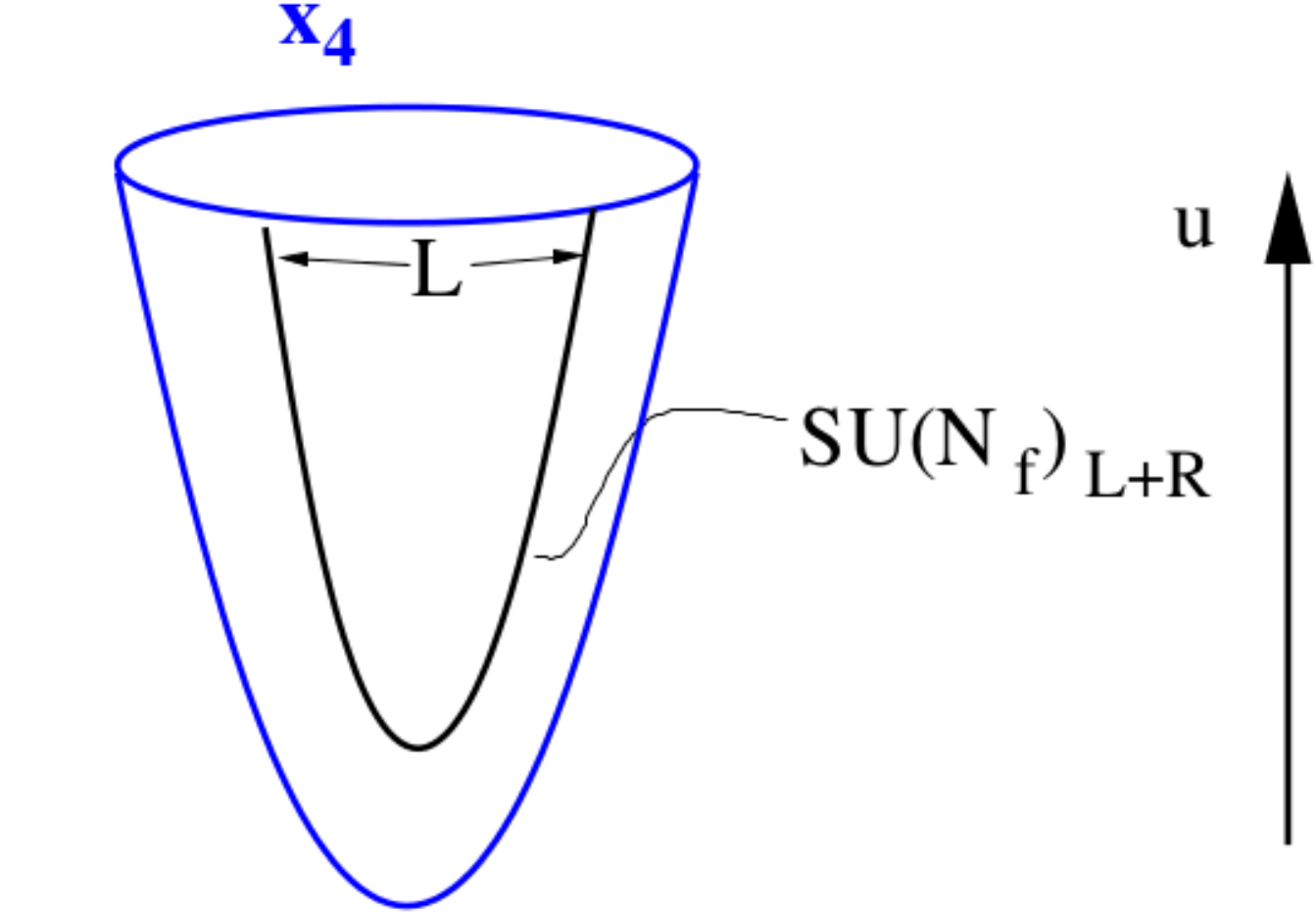}}
\caption{Left: Configuration of the stacks of $N_c$ D4 branes
and $N_f$ D8 and $\overline{\mbox{D8}}$ branes in the 10 dimensions $x_0,\ldots,x_9$,
with $x_4$ taken as periodic. Right: Geometry near the horizon of the stack of $N_c$ D4 branes, $N_c\gg N_f$, with $u$ the radial coordinate in
the transverse space $x_{5\ldots 9}$. [Figure courtesy of Andreas Schmitt]}
\label{figD4D8} 
\end{figure}

In 2004, the Witten model was extended by Sakai and Sugimoto \cite{Sakai:2004cn,Sakai:2005yt} by a configuration
of probe D8 and anti-D8 branes that fill all of the 10-dimensional space
of the Witten model except for the circle along $x^4$. This
introduces $N_f\ll N_c$ chiral quarks and antiquarks that are localized in $x^4$, but
since the D8 and anti-D8 branes have nowhere to end in the subspace spanned by $x^4$ and
the radial (holographic) direction, these branes have to connect in the bulk geometry,
leading to a simple geometric realization of chiral symmetry breaking (see Fig.~\ref{figD4D8}).

The low-energy theory of the (Witten-)Sakai-Sugimoto model involves a chiral Lagrangian for the Nambu-Goldstone
bosons with a Skyrme term, which reflects the presence of a baryon vertex provided by additional D4 branes wrapping
the $S^4$ with $N_c$ units of Ramond flux. It also provides the correct Wess-Zumino-Witten term required by
the chiral anomaly of QCD. With a minimal set of parameters, it represents arguably the most intriguing
model of holographic QCD. As reviewed in Ref.~\cite{Rebhan:2014rxa}, several features of the low-energy
spectrum of QCD are nicely reproduced.

Here we shall discuss in particular the predictions of
the Witten-Sakai-Sugimoto model for the interactions of glueballs with mesons composed of chiral quarks,
which were first studied in \cite{Hashimoto:2007ze} and revisited by us in Ref.~\cite{Brunner:2014lya,BPR}.

% \be\label{ds210}
% ds^2=\left(\frac{U}{\RSS}\right)^{3/2} \left[\eta_{\mu\nu}dx^\mu dx^\nu
% +f(U)(dx^4)^2\right]+\left(\frac{\RSS}{U}\right)^{3/2}\left[\frac{dU^2}{f(U)}+U^2 d\Omega_4^2 \right]
% \ee
% $f(U)=1-(\UKK/U)^3$

Such interactions are determined by the action of the flavor D8 and anti-D8 branes,
\bea\label{SD8full}
S_{\rm D8}&=&-T_{\rm D8}\Tr\int d^9x e^{-\Phi}
\sqrt{-\det\left(\tilde g_{MN}+2\pi\alpha' F_{MN}\right)}+S_{\rm CS}\nonumber\\
&=&-(2\pi\alpha')^2T_{\rm D8}\Tr\int d^9x e^{-\Phi}\sqrt{-\tilde g}
\left(\mathbf 1+\frac14 \tilde g^{PR}\tilde g^{QS}F_{PQ}F_{RS}+O(F^4)\right)+S_{\rm CS}
\eea
where the leading-order terms are explicitly given by
\be\label{SD8F2}
S_{\rm D8}^{(F^2)}=-\kappa\,\Tr\int d^4x \int_{-\infty}^\infty dZ\left[
\frac12 K^{-1/3}\eta^{\mu\rho}\eta^{\nu\sigma}F_{\mu\nu}F_{\rho\sigma}
+\MKK^2\eta^{\mu\nu}F_{\mu Z}F_{\nu Z}\right],\quad \kappa=\frac{\lambda N_c}{216\pi^3},
\ee
with $K\equiv 1+Z^2$, where $Z$ is a dimensionless holographic coordinate
that runs from $-\infty$ to $\infty$ as the connected branes (antipodal in $x^4$)
are followed from the boundary to the minimal value of the holographic radial coordinate $r$ and
back along the oppositely charged brane to the boundary.

The global chiral symmetry SU$(N_f)\times$SU$(N_f)$ corresponds to a
gauge symmetry with flavor gauge fields on the D8 branes. Its spontaneous breaking
is reflected by a massless mode $A_Z=\phi_0(Z)\pi(x)$ containing
the Nambu-Goldstone bosons; massive vector and axial vector mesons
are described by the even and odd eigenmodes of $A_\mu^{(n)}=\psi_n(Z) v^{(n)}_\mu(x)$.
The lowest eigenvalue gives the identification
\be
m_\rho^2=0.699314\ldots \MKK^2.
\ee
The real-world value of $m_\rho$ can therefore be used to set the overall mass scale
of the Witten-Sakai-Sugimoto model, leading to $\MKK=949$~MeV. The masses of higher vector and axial vector mesons
as predicted by the eigenvalue equation for $\psi_n(Z)$ are in fact surprisingly close
to experimental results. For example, the predicted ratios $m_{a_1}/m_\rho$ and $m_{\rho^*}/m_\rho$
deviate only by 4\% and 10\% from the measured values. This agreement may however be
somewhat fortuitous, for recent lattice simulations \cite{Bali:2013kia} at large $N_c$,
where the holographic model should actually become more reliable, lead to deviations by 21\% and 16\%, respectively.
Still, this can be viewed as a remarkable success given that the Witten-Sakai-Sugimoto model should
show deviations from large-$N_c$ QCD at scales above $\MKK$ and that already $v_\mu^{(2)}$ is
above $\MKK$.

Other predictions of the Witten-Sakai-Sugimoto model for $N_c=3$ QCD depend on the value
of the 't Hooft coupling at the scale $\MKK$ above which the model turns into 5-dimensional
super-Yang-Mills theory. 
In the original papers \cite{Sakai:2004cn,Sakai:2005yt} the pion decay constant
\be\label{fpi2}
f_\pi^2=\frac1{54\pi^4}\lambda N_c\MKK^2
\ee
was fixed to $f_\pi\approx 92.4\,{\rm MeV}$, which together with $\MKK=949$~MeV
determines the constant $\kappa$ in (\ref{SD8F2}) as
\be\label{kappaSS}
\kappa\equiv
\lambda N_c/(216\pi^3)=7.45\cdot10^{-3}
\ee
corresponding to $\lambda=16.63$.

As reviewed in Ref.~\cite{Rebhan:2014rxa}, most other possibilities of fixing $\lambda$ point to
somewhat smaller values. As an alternative we shall consider 
the recent large-$N_c$ lattice result for the ratio $m_\rho/\sqrt{\sigma}=1.504(50)$, where $\sigma$ is
the string tension, together with the holographic value
\be\label{sigmastring}
\sigma=\frac{2 \lambda}{27\pi}\MKK^2.
\ee
This gives instead $\lambda=12.55$.

Adopting the range $\lambda=16.63\ldots12.55$ for predictions of the Witten-Sakai-Sugimoto model
turns out to work remarkably well for a number of quantities of low-energy QCD at $N_c=3$.
While at infinite $N_c$, the anomaly of the $\mathrm U_A(1)$ symmetry is suppressed
and the $\eta'$ meson is a massless Goldstone boson,
at finite $N_c$ the model predicts
a finite mass through a Witten-Veneziano formula (evaluated
already in \cite{Sakai:2004cn}) with the result
\be\label{metaprime}
m_{\eta'}=\frac1{3\sqrt3 \pi}\sqrt{\frac{N_f}{N_c}}\lambda\MKK.
\ee
With $\MKK=949$~MeV and $\lambda\approx 16.63\ldots12.55$ the numerical value
for $N_c=N_f=3$ turns out to be 967\ldots730~MeV, encompassing the 
experimental value 958~MeV (while the lower value also leaves room for a contribution
due to finite quark masses that are not included in the chiral Witten-Sakai-Sugimoto model).

Another quantity that is of particular interest with regard to glueball physics
is the gluon condensate, which was calculated in Ref.~\cite{Kanitscheider:2008kd} with the result
\be\label{gluoncondensate}
C^4\equiv\left<\frac{\alpha_s}{\pi}G_{\mu\nu}^a
G^{a\mu\nu}\right>=\frac{4N_c}{3^7\pi^4}\lambda^2\MKK^4
\ee
(see Ref.~\cite{Rebhan:2014rxa} for the precise translation to the conventions of ordinary QCD).
For $\lambda= 16.63\dots12.55$ this yields $C^4=0.0126\ldots0.0072\,{\rm GeV}^4$, 
which includes the standard SVZ sum rule value \cite{Shifman:1978bx} at its upper end.

In view of the aim of using the Witten-Sakai-Sugimoto model for a calculation
of glueball decay rates, it is particularly interesting how its predictions
for the decay rates of the $\rho$ and $\omega$ meson compare with experiment.

For the effective interaction Lagrangian of $\rho$ and $\pi$ mesons one finds
\be
\mathcal L_{\rho\pi\pi}=-g_{\rho\pi\pi} 
\epsilon_{abc}(\partial_\mu \pi^a)
\rho^{b\mu}\pi^c,\quad
g_{\rho\pi\pi} =33.984
\,\lambda^{-\frac12} N_c^{-\frac12}
\ee
which yields 
\be
\Gamma_{\rho\to2\pi}/m_{\rho}=0.1535 \ldots 0.2034 \qquad \mbox{for $\lambda= 16.63\dots12.55$},
\ee
including the experimental value \cite{Agashe:2014kda} of 0.191(1).

The decay $\omega\to3\pi$, which requires the Chern-Simons part of the action (\ref{SD8F2}), was
calculated in Ref.~\cite{Sakai:2005yt}. The result is proportional to $\lambda^{-4}$
and therefore gives a larger span of predictions,
\be
\Gamma_{\omega\to3\pi}/m_{\omega}=0.0033 \ldots 0.0102 \qquad \mbox{for $\lambda= 16.63\dots12.55$}.
\ee
The experimental value is \cite{Agashe:2014kda}
0.0097(1). Encouragingly, this again fits the holographic result.

\section{Glueball masses and decay rates in the\break Witten-Sakai-Sugimoto model}

In Ref.~\cite{Hashimoto:2007ze}, the Witten-Sakai-Sugimoto models
has for the first time been used for a quantitative evaluation of
scalar glueball decay rates by including
the glueball modes in the D8 brane action. 
In Ref.~\cite{BPR}, we have
repeated, partly corrected, and extended this work.

With $\MKK$ fixed to 949~MeV through the mass of the $\rho$ meson, the
predictions of the Witten model for the glueball spectrum 
are pinned down. The result %, which does not depend on the value of $\lambda$,
is that the lowest scalar mode (\ref{deltaGG}) has the mass
$M_E=855$~MeV, which is only 10\% higher than $m_\rho$ and
thus much lower than the predictions of quenched lattice simulations.
However the lowest scalar mode in the tensor multiplet, which
is a predominantly dilatonic mode, has the mass $M_D=M_T=1487$~MeV,
not far from lattice results as far as the $0^{++}$ glueball is concerned,
while the lattice predictions for the lowest tensor glueball is about
50\% higher.

The same picture arises if one considers the ratio of glueball mass
over the square root of the string tension using Eq.~(\ref{sigmastring})
and compares with the lattice results of Ref.~\cite{Lucini:2010nv} 
obtained at large $N_c$. 
In the right panel of Fig.~\ref{figgbspctrm},
the holographic results are given by
dotted green lines for the lowest scalar (exotic) mode, which
is only half of the lattice result for the lowest $0^{++}$
state, whereas the lowest dilatonic mode (dotted black) is slightly
below the latter. (Note that ``exotic'' refers only to the polarization
of the underlying graviton mode and not to an exotic $J^{PC}$ assignment.)

Although these results do not depend on the value of $\lambda$ within
the Witten-Sakai-Sugimoto model, one
can expect important corrections involving inverse powers of $\lambda$
and $N_c$ in the form of higher-derivative terms from string-theoretic
effects beyond the supergravity approximation. However, the discrepancy
between $M_E$ and the lattice result is unusually large considering
the other predictions of the model. As already remarked, the lowest scalar mode
corresponds to an ``exotic polarization'' of the six-dimensional graviton
in that it involves metric components $\delta G_{44}$ 
referring to the extra spatial dimension whose only purpose in the Witten model
is to implement the breaking of supersymmetry.
Perhaps the exotic mode
should be discarded from predictions of the Witten-Sakai-Sugimoto model,
assuming that it will somehow disappear when the supergravity approximation
is left in the limit $\MKK\to\infty$, $\lambda\to0$ 
that brings the model towards a string-theoretic dual
of large-$N_c$ QCD.

\begin{table}
\begin{tabular}{lrc}
\hline
\tablehead{1}{l}{b}{Mode\\$J^{PC}$}  & \tablehead{1}{r}{b}{Mass\\$M$ {\rm [MeV]}}
  & \tablehead{1}{c}{b}{$\Gamma/M$\\ $(\lambda=16.63\ldots 12.55)$}
%  & \tablehead{1}{r}{b}{$\Gamma/M$\\ $(\lambda=12.55)$}   
\\
\hline
$0^{++}$ (E)\tablenote{``exotic'' polarization} & 855 & 0.092\ldots0.122 \\
$0^{++}$ & 1487 & 0.009\ldots0.012 \\
$2^{++}$ & 1487 & 0.015\ldots0.019 \\
\hline
\end{tabular}
\caption{Masses $M$ and decay rates $\Gamma_{G\to2\pi}$ of lightest scalar and tensor glueballs}
\label{tab:gbdr}
\end{table}

In Ref.~\cite{Hashimoto:2007ze}, decay rates have been studied only
for the lowest (``exotic'') scalar mode. In Ref.~\cite{BPR}, we have
also worked out the interactions of the predominantly dilatonic mode,
the tensor mode, and their excitations with pseudoscalar and vector mesons.
Writing out here only the interactions with pions, the exotic
scalar mode has an interaction Lagrangian of the form
\bea
\mathcal L^{G_E\to\pi\pi}&=&-\frac12 \left[c_1 \6_\mu\pi^a \6_\nu\pi^a
\frac{\6^\mu\6^\nu}{M_E^2}G
+\breve c_1 \6_\mu\pi^a \6^\mu\pi^a\, G\right],\nonumber\\
&&c_1=62.655\,\lambda^{-\frac12}\, N_c^{-1} \MKK^{-1},
\quad \breve c_1=16.390\,\lambda^{-\frac12}\, N_c^{-1} \MKK^{-1},
\eea
where we differ from Ref.~\cite{Hashimoto:2007ze} by a factor of $\sqrt2$
in the coefficient $c_1$ and by the term with the coefficient $\breve c_1$,
which has been dropped in Ref.~\cite{Hashimoto:2007ze}.

The predominantly dilatonic mode has an interaction with pions given by
\be
\mathcal L^{G_D\to\pi\pi}= \frac{d_1}{2} \,\partial_\mu\pi^a\partial_\nu\pi^a 
\left(\eta^{\mu\nu}-\frac{\partial^\mu \partial^\nu}{\Box}\right)D,\quad d_1=17.226\,\lambda^{-\frac12}\, N_c^{-1} \MKK^{-1},
\ee
and for the tensor mode we find
\be
\mathcal L^{G_T\to\pi\pi}= \frac{t_1}2 
\,\6_\mu\pi^a\6_\nu\pi^a \, T^{\mu\nu},\quad t_1=\sqrt6\, d_1,
\ee
where $T^{\mu\nu}$ is normalized such that
$
\mathcal L^T=\frac14 T_{\mu\nu}(\Box-M^2)
T^{\mu\nu}+B_\mu \6_\nu T^{\mu\nu}+B \eta_{\mu\nu} T^{\mu\nu} +\ldots,
$
with $B,B_\mu$ being Lagrange multiplier fields.

\begin{table}
\begin{tabular}{lrr}
\hline
decay &  $\Gamma/M$ (exp.)\tablenote{Experimental data are from the Particle Data Group \cite{Agashe:2014kda} except for those marked by a star ($\star$), which
are from Ref.~\cite{1208.0204} where the total width of $f_0(1710)$ was divided up under the assumption of
a negligible branching ratio to four pions and two $\omega$ mesons,
using data from BES \cite{hep-ex/0603048} (upper entry) and WA102 \cite{hep-ex/9907055} (lower entry),
respectively.}
  & $\Gamma/M[G_D({M^{\rm exp}})]$ \\
\hline
$f_0(1500)$ (total) & 0.072(5)  & 0.027\ldots0.037 \\
$f_0(1500)\to4\pi$ & 0.036(3)  &  0.003\ldots 0.005 \\
$f_0(1500)\to2\pi$ & 0.025(2)  & 0.009\ldots0.012\\
$f_0(1500)\to 2K$ & 0.006(1)  & 0.012\ldots0.016 \\
$f_0(1500)\to 2\eta$ & 0.004(1)  & 0.003\ldots0.004 \\
\hline
$f_0(1710)$ (total)  & 0.078(4)  &  0.059\ldots0.076 \\[4pt]
$f_0(1710)\to 2K$ &  $\star\left\{ 0.041(2) \atop 0.047(17) \right.$ & 0.012\ldots0.016 \\[12pt]
$f_0(1710)\to 2\eta$ &  $\star\left\{0.020(10) \atop 0.022(11) \right.$ & 0.003\ldots0.004 \\[12pt]
$f_0(1710)\to2\pi$ &  $\star\left\{0.017(4) \atop 0.009(2) \right.$ & 0.009\ldots0.012 \\[8pt]
$f_0(1710)\to4\pi$ & ? &  0.024\ldots 0.030 \\
$f_0(1710)\to2\omega\to6\pi$ & seen  &  0.011\ldots 0.014 \\
\hline
\end{tabular}
\caption{Experimental data for the decay rates of the isoscalar mesons $f_0(1500)$ and $f_0(1710)$
confronted with
the holographic results for various decay channels
of the predominantly dilatonic glueball ($G_D$) in the chiral limit
with glueball mass adjusted to the respective experimental values 
$M^{\rm exp}=1505$ and 1722 MeV, for 't~Hooft coupling
varied from 16.63 to 12.55.}
\label{tabexp}
\end{table}

In Table \ref{tab:gbdr} the resulting decay rates $\Gamma_{G\to2\pi}$
are given, which show that the relative width $\Gamma/M$ is small,
around 1\% and 2\%,
for the dilatonic and the tensor mode, respectively. However, the
result for the lowest (exotic) mode with mass 855 MeV is an order of
magnitude larger. (Parametrically, all three results are
of the order $\lambda^{-1}N_c^{-2}$.)
This peculiar result is in fact reminiscent of a scenario
discussed by Narison and others
in Ref.~\cite{Narison:1996fm,Narison:2005wc,Mennessier:2008kk,Kaminski:2009qg},
where both a broad glueball around 1 GeV and a heavier narrow
glueball around 1.5 GeV were found to be required by QCD sum rules. There
the lighter glueball was interpreted as a pure-glue component of the $\sigma$-meson
$f_0(500)$. Recent lattice simulations of unquenched QCD \cite{Gregory:2012hu} did not find
changes of quenched result for the mass of the lowest scalar glueball around 1.5-1.8 GeV, which
we take as motivation to concentrate on the %heavier,
predominantly dilatonic mode and to investigate how its decay rates
compare with experimental glueball candidates.

In the range of 1.5-1.8 GeV, the Particle Data Group
\cite{Agashe:2014kda} lists two isoscalar $0^{++}$ mesons, $f_0(1500)$ and $f_0(1710)$.
These are frequently
and alternatingly discussed as possible manifestations of a large
glueball component.
In Ref.~\cite{BPR} we have evaluated the decay rate into four pions
which are produced by glueball couplings with $\Tr(\rho[\pi,\pi])$ and
$\Tr(\rho\rho)$. The dilatonic mode is also heavy enough to
have decays into two kaons and two $\eta$ mesons.
In the chiral Sakai-Sugimoto model, the latter are given simply by the
flavor-symmetric factors $4/3$ and $1/3$ compared to decay into two pions.

In the upper half of Table \ref{tabexp}, the resulting decay rates
are compared with the experimental data for $f_0(1500)$
from Ref.~\cite{Agashe:2014kda}. We find that
the decay into two pions comes out about 50\% too low, while 
the decay into kaons is too large by a factor of 2. The decay
into four pions, which is the dominant decay mode of $f_0(1500)$, is
missed by an order of magnitude.

Turning to $f_0(1710)$, we need to manually adjust the mass of our
dilatonic mode in order to take into account that this meson
is above the threshold of $2\rho$ and $2\omega$. We do so by simultaneously
rescaling the mass scale in the dilaton couplings, which leaves
the dimensionless ratio $\Gamma/M$ unchanged for the decays into two
Goldstone bosons.
However the decays into four pions and also two $\omega$ mesons
(which because of their narrow width are treated as nearly stable)
now become significantly increased. Decay of $f_0(1710)$ into
two $\omega$ mesons has been seen according to the Particle Data Group
\cite{Agashe:2014kda}, but not into four pions.
The decay width into two pions [which is independent of the extrapolation
to the higher mass of the $f_0(1710)$] turns out to have the right
magnitude, but the decay into two kaons is underestimated if only
a flavor-symmetric enhancement of $4/3$ is included.

It is to be expected that a deformation of the Witten-Sakai-Sugimoto model
towards finite quark masses and thus nonzero masses for the pseudo-Goldstone
bosons such as discussed in Ref.~\cite{Hashimoto:2008sr} gives additional contributions to the vertices of
a glueball with two pseudo-Goldstone
bosons. This can significantly enhance the decays into two kaons and two
$\eta$ mesons, as suggested by the lattice results of
Ref.~\cite{Sexton:1995kd} and the analysis of Ref.~\cite{Chanowitz:2005du},
where this phenomenon was termed chiral suppression. The only feature of
our results that would then deviate from the decay pattern of $f_0(1710)$ (as
presently known) is the substantial decay rate of the holographic glueball $G_D$
into four pions. 
The experimental data for partial decay widths of $f_0(1710)$
given in Table \ref{tabexp} are from Ref.~\cite{1208.0204}, where
it was simply assumed that the partial widths for pairs of pseudoscalar mesons
add up to the total width; in the analysis of Ref.~\cite{Albaladejo:2008qa}, e.g., they
add up to only 70\%. Thus the experimental situation certainly needs further clarification,
before one can judge if and by how much the holographic result overestimates
the decay into four pions. 
\emph{If} $f_0(1710)$ is a nearly unmixed glueball as
suggested by Refs.~\cite{Lee:1999kv,Giacosa:2005zt,Cheng:2006hu,Gui:2012gx,Janowski:2014ppa},
then the (extrapolated) prediction of the Witten-Sakai-Sugimoto model
would be that there should a nonnegligible decay width into four pions and about half as much for the one into two $\omega$. (In Ref.~\cite{Janowski:2014ppa},
which employed an extended linear sigma model with a dilaton,
the decay into four pions turned out to be strongly suppressed
due to their very large value of the gluon condensate.)

In Ref.~\cite{BPR} we have also evaluated the parametrically strongly suppressed
decay of scalar glueballs into four $\pi^0$, which turns out to be significantly
smaller than the (small) rate observed for $f_0(1500)$. This also supports the 
conclusion that the result for the Witten-Sakai-Sugimoto model is not in agreement with 
a pure-glueball interpretation of $f_0(1500)$, but seems to prefer such
an interpretation for $f_0(1710)$.

For the tensor glueball we have obtained a very small decay width into two
pions. But since the lattice points to a tensor glueball far above the threshold
of two $\rho$ mesons, decays into two vector mesons (possibly also including
the $\phi$) have to be considered as well. In Ref.~\cite{BPR} we have found
that these decay channels dominate and lead to a comparatively large total width,
larger than the $f_2$ mesons listed by the Particle Data Group around and above
2 GeV, with the exception of $f_2(1950)$, which is indeed occasionally discussed
as a glueball candidate.

\section{Conclusions and outlook}

We have revisited the spectrum and the decay rates of glueballs in the Witten-Sakai-Sugimoto model which were first evaluated quantitatively in Ref.~\cite{Hashimoto:2007ze}.
At variance with the latter we have concluded that only the predominantly
dilatonic scalar glueball mode of the Witten model has a mass and decay rate comparable
with glueball candidates in the range indicated by lattice gauge theory, while
the lowest mode either has to be discarded or, more speculatively, 
could perhaps be identified with
a broader glueball component of the $\sigma$-meson (cp.~Ref.~\cite{Narison:1996fm}) 
that only drops out in the theory without quarks.

The decay pattern of the predominantly
dilatonic scalar glueball is found to deviate rather markedly from
the one observed for the glueball candidate $f_0(1500)$---its dominant decay mode into four
pions is not reproduced by the holographic results, while decay into two pions is underestimated
by a factor 2. A somewhat
better match is obtained for $f_0(1710)$, if this is assumed to be a nearly
unmixed glueball \cite{Lee:1999kv,Giacosa:2005zt,Cheng:2006hu,Gui:2012gx,Janowski:2014ppa},
as the decay rate into two pions agrees rather closely
with experimental data. The dominant decay into two kaons observed in experiments could be due to
an enhancement of glueball coupling through mass terms of pseudo-Goldstone bosons, which
will necessarily contribute to the coupling of glueballs to pseudoscalar mesons; we intend to
study the effects of deformations of the Witten-Sakai-Sugimoto model away from the chiral
limit in future work. Decay into four pions, however, are uncomfortably large when the mass of
the holographic glueball is matched to that of $f_0(1710)$, given that this decay mode has not been
observed in experiment. Decay into two $\omega$, which mostly decay into three pions, has been
observed. Our holographic result is that it should be at the level of half of the rate into four pions, and both not strongly suppressed.

At the level our analysis, no mixing of glueballs with $q\bar q$ states is taken
into account so that the results at best pertain to approximately unmixed glueballs. 
Mixing, which can strongly obscure the decay pattern of a pure glueball, is
suppressed in the Witten-Sakai-Sugimoto model by at least a factor $N_c^{-1}$ and would require
further string-theoretic input to be studied semi-quantitatively. Absent that, it might be interesting to incorporate our holographic
results in phenomenological models of mixing between glueballs and $q\bar q$ states.

%%%%%%%%%%%%%%%%%%%%%%%%%%%%%%%%%%%%%%%%%%%%%%%%

\begin{theacknowledgments}We would like to thank 
Koji Hashimoto, Chung-I Tan, and Seiji Terashima for correspondence and
David Bugg, Francesco Giacosa, Stanislaus Janowski, and Dirk Rischke for useful discussions.
This work was supported by the Austrian Science
Fund FWF, project no. P26366, and the FWF doctoral program
Particles \& Interactions, project no. W1252.
\end{theacknowledgments}



\bibliographystyle{aipproc}   % if natbib is available

\bibliography{hgdr}

\IfFileExists{\jobname.bbl}{}
 {\typeout{}
  \typeout{******************************************}
  \typeout{** Please run "bibtex \jobname" to optain}
  \typeout{** the bibliography and then re-run LaTeX}
  \typeout{** twice to fix the references!}
  \typeout{******************************************}
  \typeout{}
 }

\end{document}